\def\rfr#1{eq. (\ref{#1})}

\def\dert#1#2{\frac{{{d}}{#1}}{{{d}}{#2}}}              

\def\virg#1{``#1''}

\def\eqi{\begin{equation}}
\def\eqf{\end{equation}}
\def\eqia{\begin{eqnarray}}
\def\eqfa{\end{eqnarray}}
\def\rp#1#2{{#1\over#2}} \def\lb#1{\label{#1}}

\def\bds#1{\boldsymbol{#1}}

\documentclass[usenatbib]{mn2e}
\usepackage{amsmath}
\usepackage{amscd}
\usepackage{amssymb}
\usepackage{latexsym}
\usepackage{graphicx}
\usepackage{epsfig}

\title[The Neptunian system  and the Pioneer anomaly]{Does the Neptunian system of satellites challenge a gravitational origin for the Pioneer anomaly?}

\author[L. Iorio]{
L. Iorio$^{1}$\thanks{E-mail:
lorenzo.iorio@libero.it}\\
$^{1}$INFN-Sezione di Pisa, Viale Unit$\grave{\rm a}$ di Italia 68, 70125, Bari (BA), Italy
}

\begin{document}



\maketitle

\label{firstpage}

%
%
%
%
%
%
%
%
%
%
%
%
\begin{abstract}
If the Pioneer Anomaly (PA) was a genuine dynamical effect of gravitational origin, it should also affect the orbital motions of the solar system's bodies moving in the space regions in which the PA manifested itself in its presently known form, i.e. as a constant and uniform acceleration approximately directed towards the Sun with a non-zero magnitude
$A^{\rm Pio} = (8.74\pm 1.33)\times 10^{-10}\ {\rm m\ s}^{-2}$ after 20 au from the Sun. In this paper we preliminarily investigate its effects on the orbital motions of the Neptunian satellites Triton, Nereid and Proteus, located at about 30 au from the Sun, both analytically and numerically.   Extensive observational records covering several orbital revolutions have recently been analyzed for them, notably improving the knowledge of their orbits. { Both} analytical { and numerical} calculations, limited to the direct, Neptune-satellite interaction, show that { the peak-to-peak amplitudes of the} PA-induced { radial, transverse and out-of-plane perturbations over one century are up to  300 km, 600 km, 8 m for Triton, $17,500$ km, $35,000$ km, 800 km for Nereid, and 60 km, 120 km, 30 m for Proteus.}
The corresponding orbital uncertainties obtained from a recent analysis of all the data available for the satellites considered are, in general, smaller { by one-two orders of magnitude}, although obtained without modeling a Pioneer-like extra-force.
Further investigations based on a re-processing of the satellites'{ real or simulated} data with modified equations of motions including an additional Pioneer-type force as well are worth  being implemented and may shed further light on this important issue.
\end{abstract}



\section{Introduction}
The Pioneer Anomaly (PA)  \citep{Nie06} consists of an unmodeled, almost constant and uniform
acceleration approximately directed towards the Sun and of magnitude \citep{And98,And02a}
  \eqi A^{\rm Pio} = (8.74\pm 1.33)\times 10^{-10}\ {\rm m\ s}^{-2}.\lb{pioa}\eqf
It was  detected in the radiometric data from the Pioneer 10/11
spacecraft after they passed the 20
au threshold moving along roughly antiparallel
escape hyperbolic paths taken after their previous encounters with Jupiter ($\sim 5$ au)
and Saturn ($\sim 10$ au), respectively. The PA's existence has been subsequently confirmed by independent investigations
by \cite{Mar02}, \citet{Ols07} and \citet{Levy} as well. Interestingly, latest data analyses are focussing on periodic variations
of the anomaly, characterized as functions of the azimuthal angle $\varphi$
defined by the directions Sun-Earth antenna and Sun-Pioneer \citep{Levy}.
Concerning the possibility that it started
to manifest itself at shorter heliocentric distances \citep{NI1,NI2}, efforts to retrieve and analyze
early data from Pioneer 10/11 are currently being made \citep{reco1,reco2}.
The PA is one of some astrometric anomalies in the solar system  reported in recent years \citep{Lam08,And09,Ior09}.

 Attempts to explain some features of the PA in terms of mundane, non-gravitational effects, pertaining the Pioneer probes themselves like thermal forces among different parts of the spacecraft \citep{And02a,Mur99,Kat99,Sche03,Mbe04,Berto08,Tot09} or external influences like anisotropic solar emission \citep{Mash},  have been undertaken, but some of them  have not obtained full consensus so far \citep{And99a,And99b,Mash}.
 On the other hand, latest work by \citet{Berto08} strongly points out that PA is a thermal effect due to the energy sources in the spacecraft; further studies on possible thermal effects like a potential asymmetric heat dissipation of the spacecraft
surface are ongoing \citep{Riv}.
 Conventional explanations of gravitational origin in terms of drag due to interplanetary dust, dark matter, Kuiper Belt Objects \citep{And02a,Nie05,Nieto05,deDie,Berto06} have been found not satisfactorily as well. As a consequence, many suggestions invoking non-standard gravitational and non-gravitational  physics like non-linear electrodynamics \citep{nonlin} have been proposed. For a review see, e.g., \citep{And02a,Dit05,Bert06,Izzo,Diego} and references therein. Among the various proposed exotic gravitational mechanisms we recall those by \citet{mof}, based on a long-range Yukawa-like extra-force, and by \citet{jek} who proposed metric extensions of the Einstein's General Theory of Relativity (GTR).
 Attempts to find exotic gravitational explanations for PA did not even cease after the publication of the latest works on the non-gravitational effects like \citet{Berto08}; just to limiting to published works, see, e.g. \citet{Avra,Wil,Grea,Exi}. A dedicated spaceraft-based mission to test the PA in the outer regions of the solar system has also been proposed and investigated \citep{Dit05,Izzo,Bert07}.

The hypothesis that non-standard forces of gravitational origin are able to explain the anomalous behavior of the Pioneer spacecraft must cope with the following crucial remark. If the PA
was due to some modifications of the known laws of gravity, this should be due to a
radial extra-force affecting the orbits of the astronomical bodies (planets and their satellites, comets, Trans-Neptunian Objects, etc.) as well, especially those moving
in the space regions in which  the PA manifested itself in its presently known form. Otherwise, a violation of the equivalence principle largely incompatible with the present-day bounds of $\sim 10^{-13}$ from Earth-based laboratory experiments \citep{equi} would occur. The
impact of a Pioneer-like additional acceleration on the motion of planets and minor
bodies in the outer regions of the solar system interested by the PA was recently studied by numerous
authors with different approaches \citep{And02b,Izzo,Iorio06,Page06,Pit06,Iorio07a,Iorio07b,Tan07,Wal07,Sta08,Iorio09,Page09,Fie09}.
In particular, \citet{And02b} discussed
the impact of a Pioneer-like acceleration on the long-period comets and  the form of the Oort cloud; however, such bodies are not particularly well suited to perform accurate tests of gravitational theories because of the impact of several aliasing non-gravitational perturbations like out-gassing  as they approach the
Sun. \citet{Page06} investigated the potential offered by an analysis of the minor planets in the outer solar system to confirm or refute the existence of a gravitational Pioneer effect. \citet{Wal07} used a well-observed sample of Trans-Neptunian Objects (TNOs) between 20 and 100 au from the Sun to constrain Pioneer-like deviations from Newtonian gravity in that region of the solar system. By fitting the TNOs' observations with modified equations of motion according to \rfr{pioa}, \citet{Wal07} found $(0.87\pm 1.6)\times 10^{-10}$ m s$^{-2}$, which is consistent with zero and whose upper bound is inconsistent with \rfr{pioa} at $4\sigma$ level.
\citet{Izzo}, \citet{Iorio06}, \citet{Iorio07a} and \citet{Tan07} looked at the outer planets. \citet{Izzo} parameterized the PA in terms of a change of the effective
reduced solar mass felt by Neptune finding it nearly two
orders of magnitude beyond the current observational constraint. Moreover, they noted that the Pioneer 11 data contradict the Uranus ephemerides-obtained without explicitly modeling the PA-by more than one order of magnitude. \citet{Iorio06}, \citet{Iorio07a} and \citet{Tan07} computed the secular effects induced by an unform and radial extra-acceleration like that of \rfr{pioa} on the orbits of Uranus, Neptune and Pluto, located at 20-40 au from the Sun, and compared them to the present-day, unmodified  ephemerides. \citet{Iorio06} and \citet{Iorio07a} concluded that the resulting anomalous effects on all of them would be too large to have escaped  from detection so far. Doubts concerning Neptune were raised by \citet{Tan07} in the sense that the accuracy of the currently available observations for it would not, in fact, exclude the possibility that Neptune is acted upon by $A_{\rm Pio}$. Other authors made a step further by including a Pioneer-like extra-acceleration in the force models  and fitting again the planetary observations with such modified equations of motion. More specifically, \citet{Page09} fitted modified dynamical models including \rfr{pioa} to observational records for Uranus, Neptune and Pluto showing that the current ephemeris of Pluto does not preclude the existence of the Pioneer effect because its orbit would not be well enough characterized at present to make such an assertion. \citet{Sta08} fitted planetary data records with a modified version of the JPL DE ephemerides with a uniform extra-acceleration directed towards the Sun acting on Uranus, Neptune and Pluto; a magnitude as small as just $10\%$ of \rfr{pioa} yielded completely unacceptable residuals for all the three outer planets.  \citet{Fie09} added an extra-acceleration like that of \rfr{pioa} to the equations of motion of the outer planets and fitted the resulting modified ephemerides to their observations by finding that Uranus excludes the existence of  Pioneer-like acceleration as large as \rfr{pioa} at a $4\sigma$ level. On the contrary, for Neptune and Pluto
the effect of \rfr{pioa} is absorbed by the fit, so that the resulting residuals do not allow to exclude the existence of a Pioneer-like anomalous acceleration affecting such bodies.  The existence of a standard PA in the regions crossed by Jupiter and Saturn has been ruled out by \citet{Iorio07b} and \citet{Sta08} with different approaches. For non-standard, velocity-dependent forms of the PA and their compatibility with different ephemerides of the outer planets, see \citet{Sta08,Sta10} and \citet{Iorio09}; { such different approaches show that almost all of them  are not compatible with the planetary observations}.

In this paper we investigate a different astronomical laboratory with respect to those examined so far to put on the test the hypothesis of the gravitational origin of the PA independently of what detected in the Pioneer 10/11 telemetry. Indeed, we will look at the orbital effects induced by a Pioneer-like acceleration directed towards the Sun on the Neptunian satellites Triton, Nereid and Proteus in view of the recent improvements in their orbit determination \citep{Jac09} based on the analysis of extensive data records covering several orbital revolutions. Their Keplerian orbital elements are listed in Table \ref{satelem}.
\begin{table}
\caption{Semimajor axis $a$, eccentricity $e$, inclination $I$, longitude of the ascending node $\Omega$, argument of pericenter $\omega$ and orbital period $P_{\rm b}$ of Triton, Nereid and Proteus. Reference frame: ICRF/J2000.0  with Neptune as center body. Reference system: Ecliptic and Mean Equinox of Reference Epoch. Epoch: 31 October 1989. The HORIZON interface by NASA has been used. The orbital period of Neptune is $164.9$ yr.}
\label{satelem}
\begin{tabular}{llll}\hline
Keplerian orbital element & Triton & Nereid  & Proteus  \\
\hline
$a$ (km) & $354,767$ & $5,517,147$ & $117,714$\\
$e$ & $0.00003$ & $0.75428$ & $0.00090$\\
$I$ (deg) & $130.9$ & $5.0$ & $28.9$\\
$\Omega$ (deg) & $213.2$ & $320.3$ & $48.1$\\
$\omega$ (deg) & $60.2$ & $296.1$ & $54.4$\\
$P_{\rm b}$ (d) & $5.8$ & $360.4$ & $1.1$ \\
\hline
\end{tabular}
\end{table}
In this paper we will perform a preliminary sensitivity analysis by means of analytical and numerical calculations. Their goal is to
check if the scenario considered is worth  further, more detailed investigations. They could involve, e.g., a re-processing of the Neptunian satellites' { real or simulated} data sets
with modified equations of motion including a standard gravitational Pioneer-like extra-acceleration radially directed towards the Sun as well.

In Section \ref{secdue} we first analytically work out the anomalous PA-type orbital effects on Triton, Nereid and Proteus (Section \ref{subsec1}). Then, we perform numerical integrations of their equations of motion with and without the PA. Finally, we compare our results to the latest determinations of the orbital accuracies for such satellites (Section \ref{subsec2}). Section \ref{conclu} is devoted to the conclusions.
\section{Effects of a standard   Pioneer Anomaly on the Neptune's satellites}\lb{secdue}
We  use ICRF/J2000.0  with Neptune as center body as reference frame. To be consistent with \citet{Jac09}, we adopt 31 October 1989 as reference epoch. The reference system used has the ecliptic and mean equinox of reference epoch. In such a frame a standard Pioneer-like acceleration
has the form
\eqi\bds A^{\rm Pio} = A^{\rm Pio}\bds n_{\odot},\eqf
where $\bds n_{\odot}$ is the unit vector pointing towards the Sun displayed  in Table \ref{SUN_coord}.
\begin{table}
\caption{Cartesian coordinates and distance of the Sun, in au, at the epoch. Reference frame: ICRF/J2000.0  with Neptune as center body. Reference system: Ecliptic and Mean Equinox of Reference Epoch. Epoch: 31 October 1989. The HORIZON interface by NASA has been used.}
\label{SUN_coord}
\begin{tabular}{llll}\hline
$x_{\odot}$ (au) & $y_{\odot}$ (au) & $z_{\odot}$ (au) & $r_{\odot}$ (au)\\
\hline
$-6.2$  & $29.5$  & $-0.4$ & $30.1$\\
\hline
\end{tabular}
\end{table}
As a consequence, $\bds A^{\rm Pio}$ has the components shown in Table \ref{Pio_comp} at the reference epoch.
\begin{table}
\caption{Components of $\bds A^{\rm Pio}= A^{\rm Pio}\bds n_{\odot}=A^{\rm Pio}/r_{\odot}\{x_{\odot},y_{\odot},z_{\odot}\}$, in m s$^{-2}$, at the epoch. Reference frame: ICRF/J2000.0  with Neptune as center body. Reference system: Ecliptic and Mean Equinox of Reference Epoch. Epoch: 31 October 1989.}
\label{Pio_comp}
\begin{tabular}{lll}\hline
$A_x^{\rm Pio}$ (m s$^{-2}$) & $A_y^{\rm Pio}$ (m s$^{-2}$) & $A_z^{\rm Pio}$ (m s$^{-2}$)\\
\hline
$-1.79\times 10^{-10}$  & $8.55\times 10^{-10}$  & $-0.13\times 10^{-10}$  \\
\hline
\end{tabular}
\end{table}
It can be noted that it is mainly directed along the $y$ axis of the chosen frame.
Since we are interested in its secular, i.e. averaged over one orbital period, effects on the motion of the Neptunian satellites, we can safely consider
$\bds A^{\rm Pio}$ as constant because of the short satellites' periods (see Table \ref{satelem}) with respect to the Neptunian one amounting to $164.9$ yr. In other words, each satellite faces the action of a constant and uniform disturbing acceleration directed along a generic direction in space which, in general, does not coincide with the Neptune-satellite radial one.
\subsection{Analytical and numerical calculation}\lb{subsec1}
The orbital effects of such an anomalous acceleration can be worked out with standard perturbative techniques by using, e.g., the Gauss equations for the variations of the elements \citep{Ber}
{{
\begin{eqnarray}\lb{Gauss}
\dert a t & = & \rp{2}{n\eta} \left[e A_R\sin f +A_{T}\left(\rp{p}{r}\right)\right],\lb{gaus_a}\\  \nonumber \\
\dert e t  & = & \rp{\eta}{na}\left\{A_R\sin f + A_{T}\left[\cos f + \rp{1}{e}\left(1 - \rp{r}{a}\right)\right]\right\},\lb{gaus_e}\\ \nonumber \\
\dert I t & = & \rp{1}{na\eta}A_N\left(\rp{r}{a}\right)\cos u,\\  \nonumber \\
\dert\Omega t & = & \rp{1}{na\sin I\eta}A_N\left(\rp{r}{a}\right)\sin u,\lb{gaus_O}\\   \nonumber \\
\dert\omega t & = &\rp{\eta}{nae}\left[-A_R\cos f + A_{T}\left(1+\rp{r}{p}\right)\sin f\right]-\cos I\dert\Omega t,\lb{gaus_o}\\  \nonumber \\
\dert {\mathcal{M}} t & = & n - \rp{2}{na} A_R\left(\rp{r}{a}\right) -\eta\left(\dert\omega t + \cos I \dert\Omega t\right),\lb{gaus_M}
\end{eqnarray}
}}
where  ${\mathcal{M}}$ is the mean anomaly of the orbit of the test particle,  $f$ is its true anomaly reckoned from the pericentre position, $u\doteq\omega+f$ is the argument of latitude, $n\doteq\sqrt{GM/a^3}=2\pi/P_{\rm b}$ is the unperturbed Keplerian mean motion ($G$ is the Newtonian constant of gravitation and $M$ is the mass of the central body), $\eta\doteq\sqrt{1-e^2}$ and $p\doteq a(1-e^2)$ is the semi-latus rectum. $A_R,A_T,A_N$ are the projections of the perturbing acceleration $\bds A$ onto the radial $R$, transverse $T$ and out-of-plane $N$ directions of the particle's co-moving frame whose time-varying unit vectors are \citep{Mont}
\eqi \hat{{\bds r}} =\left(
       \begin{array}{c}
          \cos\Omega\cos u\ -\cos I\sin\Omega\sin u\\
          \sin\Omega\cos u + \cos I\cos\Omega\sin u\\
         \sin I\sin u \\
       \end{array}
     \right)
\eqf
 \eqi \hat{{\bds t}} =\left(
       \begin{array}{c}
         -\sin u\cos\Omega-\cos I\sin\Omega\cos u \\
         -\sin\Omega\sin u+\cos I\cos\Omega\cos u \\
         \sin I\cos u \\
       \end{array}
     \right)
\eqf
 \eqi \hat{{\bds n}} =\left(
       \begin{array}{c}
          \sin I\sin\Omega \\
         -\sin I\cos\Omega \\
         \cos I\\
       \end{array}
     \right)
\eqf
A straightforward calculation shows that the $R-T-N$ components of a constant and uniform perturbing acceleration, like our $\bds A^{\rm Pio}$ over the timescales involved here, are linear combinations
of $A_x$, $A_y$, $A_z$ with coefficients proportional to
harmonic functions whose arguments are, in turn, linear combinations
of $u$, $\Omega$ and $I$.Thus, in the satellite's co-moving frame $\bds A^{\rm Pio}$ is time-dependent through $f$ in $u$. In order to have the secular perturbations of the Keplerian orbital elements, $A_R,A_T,A_N$ have to be inserted into the right-hand-sides of the Gauss equations which must be evaluated onto the unperturbed Keplerian ellipse
\eqi r = a(1-e\cos E),\eqf where $E$ is the eccentric anomaly, and integrated over a full orbital revolution by means of
\eqi \rp{dt}{P_{\rm b}} = \rp{(1-e\cos E)}{2\pi}dE.\eqf Other useful relations are
\eqi \cos f = \rp{\cos E -e}{1-e\cos E},\ \sin f = \rp{\sqrt{1-e^2}\sin E}{1-e\cos E}.\eqf
After cumbersome calculations it turns out that, { apart from the semimajor axis $a$ whose secular rate vanishes}, all the other Keplerian orbital elements $\psi$ experience non-vanishing secular precessions of the form\footnote{ The semimajor axis $a$ does not appear in the denominator of the equation for the eccentricity rate.}
\eqi \left\langle\dot\psi\right\rangle=\frac{\mathcal{C}^{(\psi)}_x A_x + \mathcal{C}^{(\psi)}_y A_y +\mathcal{C}^{(\psi)}_z A_z}{na},\ \psi = I,\Omega,\omega,\mathcal{M}.\lb{lunga}\eqf
In it,
\eqi \mathcal{C}^{(\psi)}_{j}=\sum_{k}F^{(\psi)}_{jk}(e)\cos\xi^{(\psi)}_{jk},\ j=x,y,z,\eqf
where
$F^{(\psi)}_{jk}(e)$ are complicated functions of the eccentricity
and $\xi^{(\psi)}_{jk}$ are linear combinations of the longitude of the pericenter $\varpi\doteq\omega+\Omega$, $\Omega$ and $I$.
In principle,
they are time-varying according to
\begin{eqnarray}
\varpi = \varpi_0+\dot\varpi t,\\
\Omega = \Omega_0+\dot\Omega t,\\
I = I_0+\dot I t;
\end{eqnarray}
from a practical point of view, since their secular rates
 are quite smaller, especially for Triton and Nereid, we can assume $\varpi\approx\varpi_0$, $\Omega\approx\Omega_0$, $I\approx I_0$ in computing $\cos\xi_{jk}$, where $\varpi$, $\Omega_0$, $I_0$ are their
values at  epoch (see Table \ref{satelem}).

The $R-T-N$  shifts over a generic time interval $\Delta t$  can be exactly computed according to
{
\citet{Cas}. The radial perturbation
is
\eqi\Delta R=K_a\Delta a+K_e\Delta e+K_{\mathcal{M}}\Delta\mathcal{M},\lb{radiale}\eqf where
\begin{eqnarray}
   K_a &=& \rp{r}{a}, \\
   K_e &=& -a\cos f, \\
  K_{\mathcal{M}}&=& \rp{ae}{\sqrt{1-e^2}}\sin f.
\end{eqnarray}
\rfr{radiale} shows that it would be incorrect to identify the shift in the radial component of the orbit with
the perturbation of the semimajor axis only. Otherwise, misleading conclusions concerning the mean motion $n$ and, thus, the transverse component as well could be traced. Indeed, if, say, a secular signature in $\Delta R$ was found, from the identification $\Delta R=\Delta a$ it could be argued that an analogous perturbation in the mean motion \eqi\rp{\Delta n}{n}=-\rp{3}{2}\rp{\Delta a}{a}\eqf would occur as well. As a consequence, a quadratic effect in the transverse component should occur through the perturbed mean longitude. Actually, this does not happen in our case: indeed, we will show that, although, as already noted, no secular effects on $a$ are present, both the radial and the transverse components exhibit  cumulative perturbations with secular trends, without any quadratic signature in the transverse one.
The transverse perturbation is
\eqi\Delta T=H_e\Delta e+H_{\mathcal{M}}\Delta{\mathcal{M}}+r(\Delta\omega+\cos I\Delta\Omega),\lb{trave}\eqf
with
\begin{eqnarray}
   H_e &=& a\left(1+\rp{1}{1-e^2}\rp{r}{a}\right)\sin f, \\
  H_{\mathcal{M}}&=& \rp{a^2\sqrt{1-e^2}}{r}.
\end{eqnarray}
The out-of-plane perturbation is
\eqi\Delta N =r\left(\Delta I\sin u-\Delta\Omega\sin I\cos u\right).\lb{outof}\eqf
In \rfr{radiale}, \rfr{trave} and \rfr{outof} the perturbations of the Keplerian orbital elements have to be intended as
\eqi\Delta\psi=\int_0^{E}d\psi,\ \psi=a,e,I,\Omega,\omega,\mathcal{M},\eqf
where $d\psi$ is taken from \rfr{gaus_a}-\rfr{gaus_M}.
The explicit expressions of $\Delta R^{\rm Pio},\Delta T^{\rm Pio}, \Delta N^{\rm Pio}$  are rather cumbersome, so that we will not explicitly show them.
  It turns out that linearly growing signatures are present in all of them along with sinusoidal terms; quadratic terms are, instead, absent.

In Figure \ref{Triton_anal}-Figure \ref{Proteus_anal} we plot the Pioneer-induced $R-T-N$ perturbations for Triton, Nereid and Proteus over a century. In order to express the eccentric anomaly as a function of time we used a partial sum of the series\footnote{It converges for all $e<1$. See on the WEB: http://mathworld.wolfram.com/KeplersEquation.html}
\eqi E=\mathcal{M}+2\sum_{s=1}^{\mathcal{N}}\rp{J_s(se)}{s}\sin(s\mathcal{M}),\ \mathcal{M}\doteq n (t-t_0),\lb{kepeq}\eqf
where $J_s(se)$ are the Bessel functions of the first kind.
\begin{figure}
\centerline{
\vbox{
\epsfysize= 5.5 cm\epsfbox{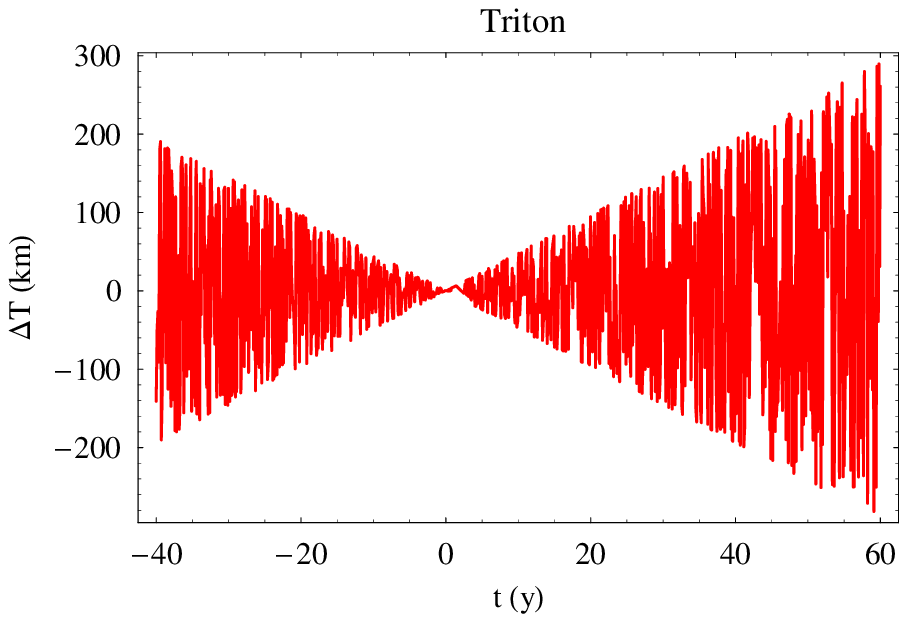}
\epsfysize= 5.5 cm\epsfbox{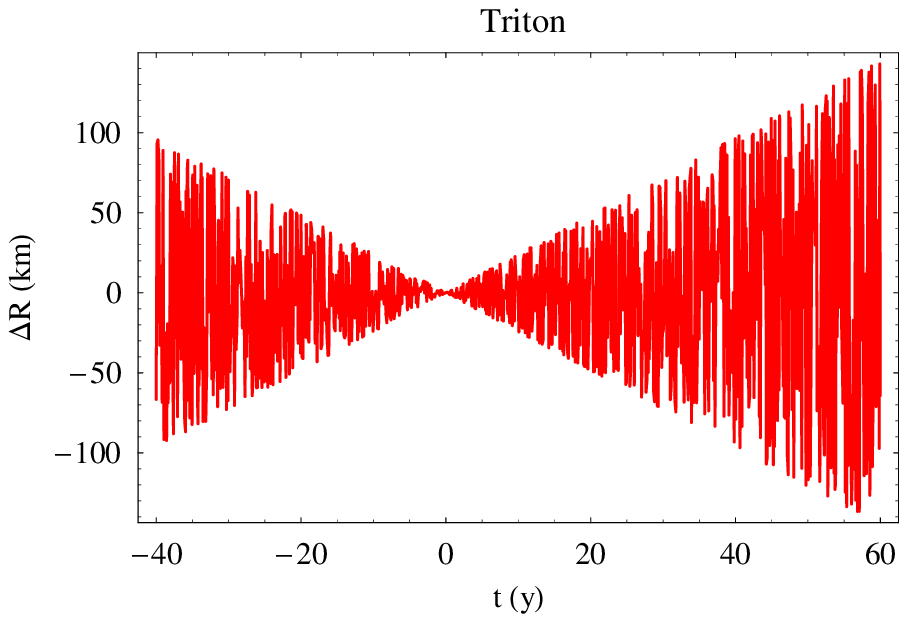}
\epsfysize= 5.5 cm\epsfbox{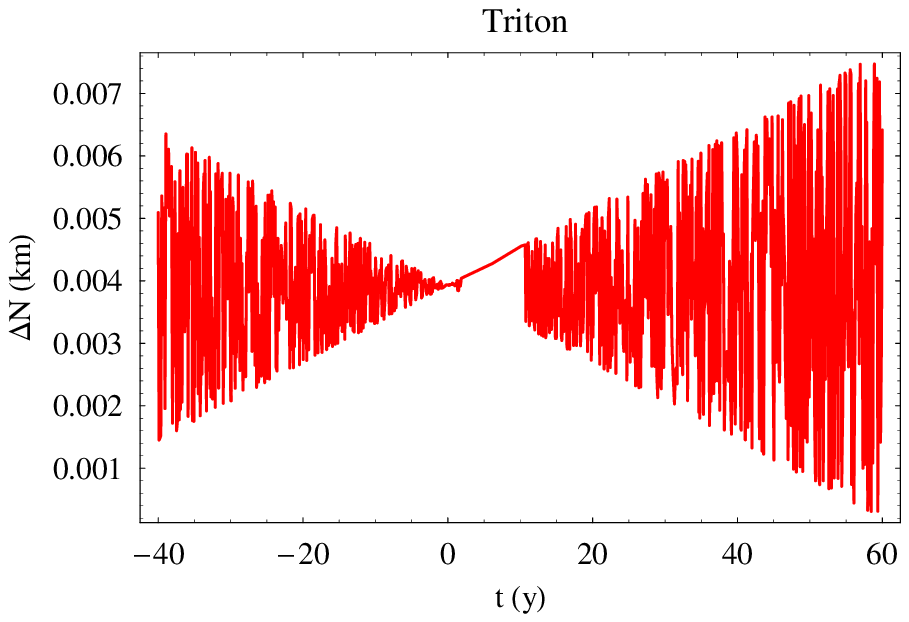}
      }
}
\caption{Analytically computed transverse, radial and out-of-plane Pioneer-induced shifts for Triton over 100 yr according to \rfr{radiale}, \rfr{trave} and \rfr{outof}, and Table \ref{satelem}. In the expansion of \rfr{kepeq} we retained just the first term because, in view of the extremely small eccentricity of the orbit of Triton, the second term is of the order of $10^{-10}$.\label{Triton_anal}}
\end{figure}
\begin{figure}
\centerline{
\vbox{
\epsfysize= 5.5 cm\epsfbox{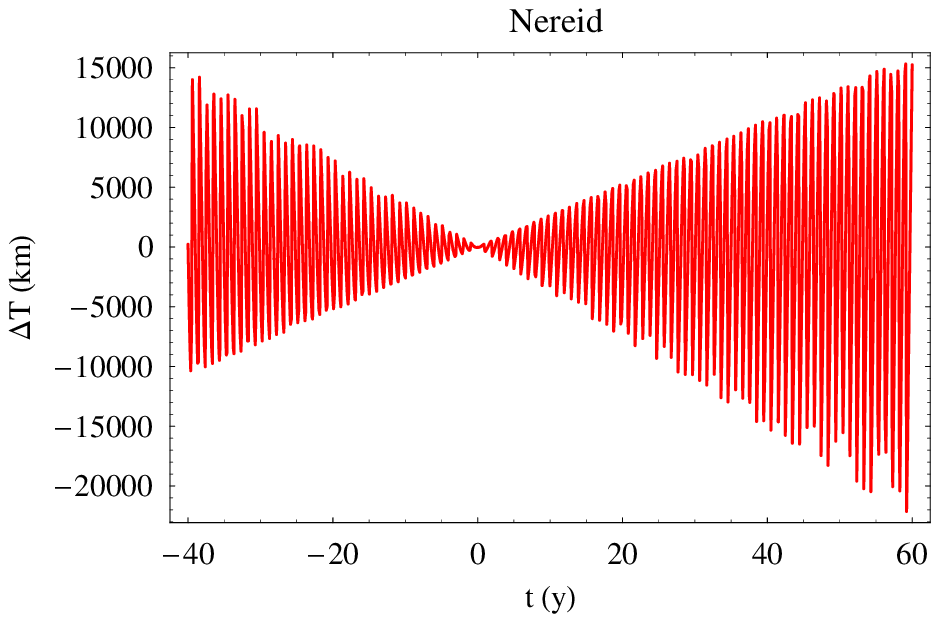}
\epsfysize= 5.5 cm\epsfbox{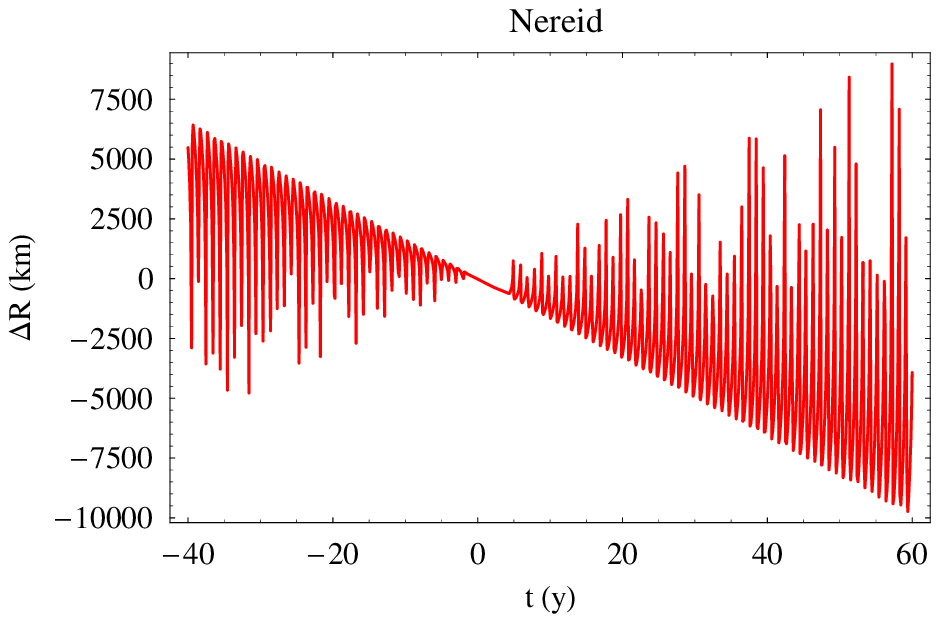}
\epsfysize= 5.5 cm\epsfbox{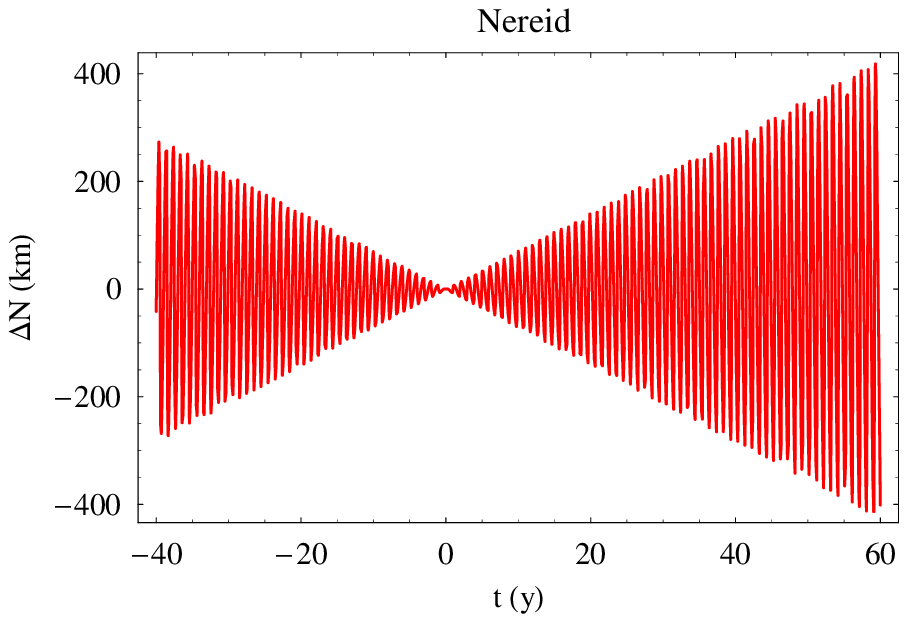}
      }
}
\caption{Analytically computed transverse, radial and out-of-plane Pioneer-induced shifts for Nereid over 100 yr according to \rfr{radiale}, \rfr{trave} and \rfr{outof}, and Table \ref{satelem}. In the expansion of \rfr{kepeq} we retained the first 40 terms because, in view of the  large eccentricity of the orbit of Triton, the following ones are of the order of, or smaller than $10^{-6}$.\label{Nereid_anal}}
\end{figure}
\begin{figure}
\centerline{
\vbox{
\epsfysize= 5.5 cm\epsfbox{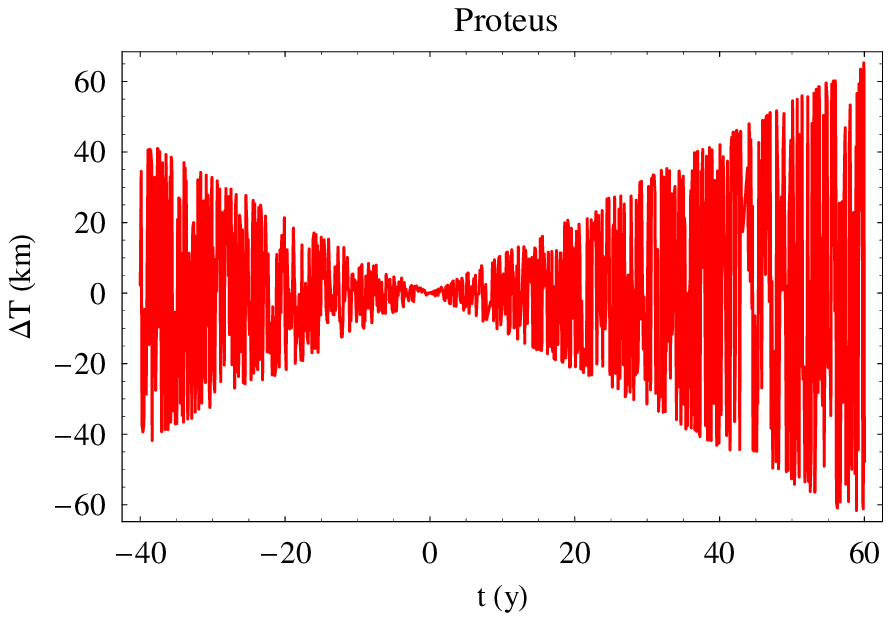}
\epsfysize= 5.5 cm\epsfbox{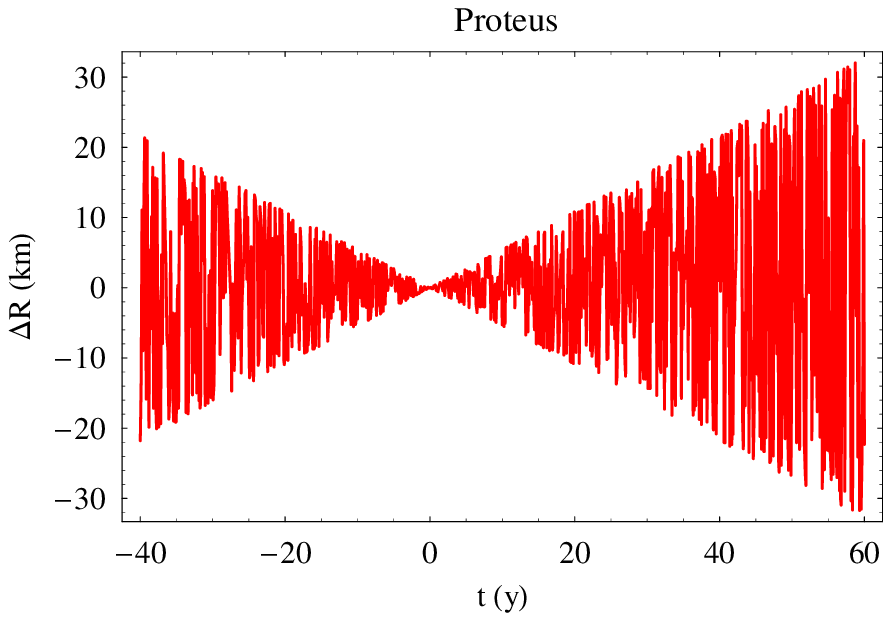}
\epsfysize= 5.5 cm\epsfbox{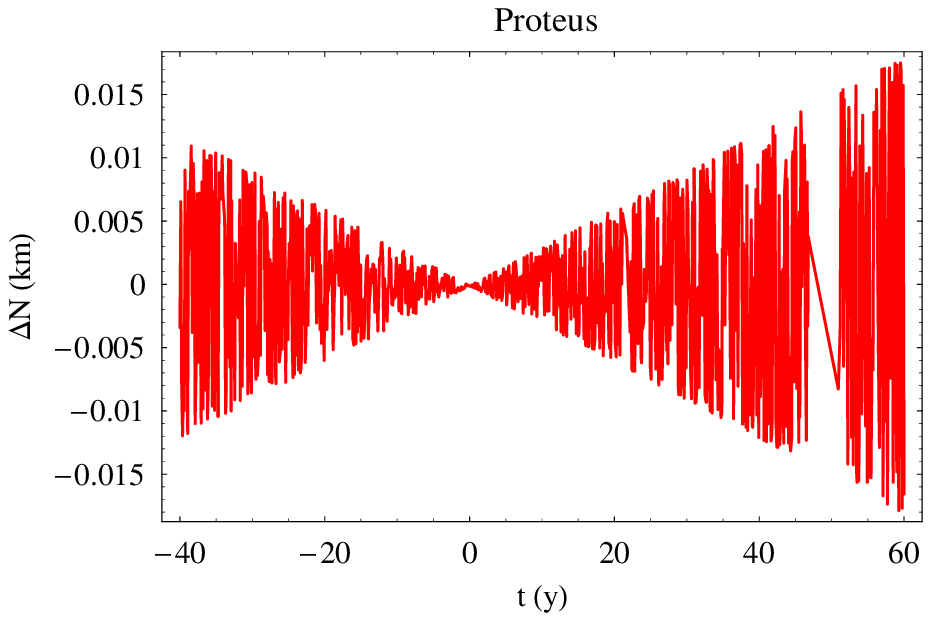}
      }
}
\caption{Analytically computed transverse, radial and out-of-plane Pioneer-induced shifts for Proteus over 100 yr according to \rfr{radiale}, \rfr{trave} and \rfr{outof}, and Table \ref{satelem}. In the expansion of \rfr{kepeq} we retained just the first term because, in view of the extremely small eccentricity of the orbit of Triton, the second term is of the order of  $10^{-7}$.\label{Proteus_anal}}
\end{figure}

The peak-to-peak
amplitudes for the Pioneer-type $R-T-N$ perturbations are  300 km, 600 km, 8 m for Triton, $17,500$ km, $35,000$ km, 800 km for Nereid, and 60 km, 120 km, 30 m for Proteus.
}


%
%
%
%
%
%
%
It should, now, be pointed out that we have only considered the direct perturbations induced by $\bds A^{\rm Pio}$ on each satellite considered separately; in fact, the total effect may be even larger because of the mutual gravitational interactions among the satellites themselves and Uranus which are all allegedly influenced by the PA as well. Moreover, it can be argued that, over time intervals larger than one orbital period { as those used here}, the Pioneer-induced signatures are modulated by the slowly changing $\Omega,\omega, I$
 because of Neptune's oblateness and N$-$body interactions with the other giant planets and satellites themselves, { and by the variation of the Sun's position which reflects into slow changes in the components of $\bds A^{\rm Pio}$}.

{ To support our analytical calculation and to further clarify the issue of the semimajor axis}, we also performed  numerical integrations with MATHEMATICA of the equations of motion of Triton, Nereid and Proteus with and without the PA. Concerning the classical forces common to all the three satellites, we included the first two even zonal harmonics $J_2, J_4$ of Neptune and the attraction of Uranus, Saturn and Jupiter. The intersatellite interactions have been taken into account as well by considering Nereid and Proteus as massless point particles acted upon by a massive Triton, as done by \cite{Jac09}.
{ First, in Figure \ref{semia} we plot the Pioneer-type perturbations on the semimajor axes $a$ of Triton, Nereid and Proteus over one Keplerian orbital period. As expected from our analytical calculation, no cumulative, net effects occur; this is neither in contradiction with the absence of a quadratic signature in $\Delta T$ nor with the presence of a linear signature in $\Delta R$.
\begin{figure}
\centerline{
\vbox{
\epsfysize= 5.5 cm\epsfbox{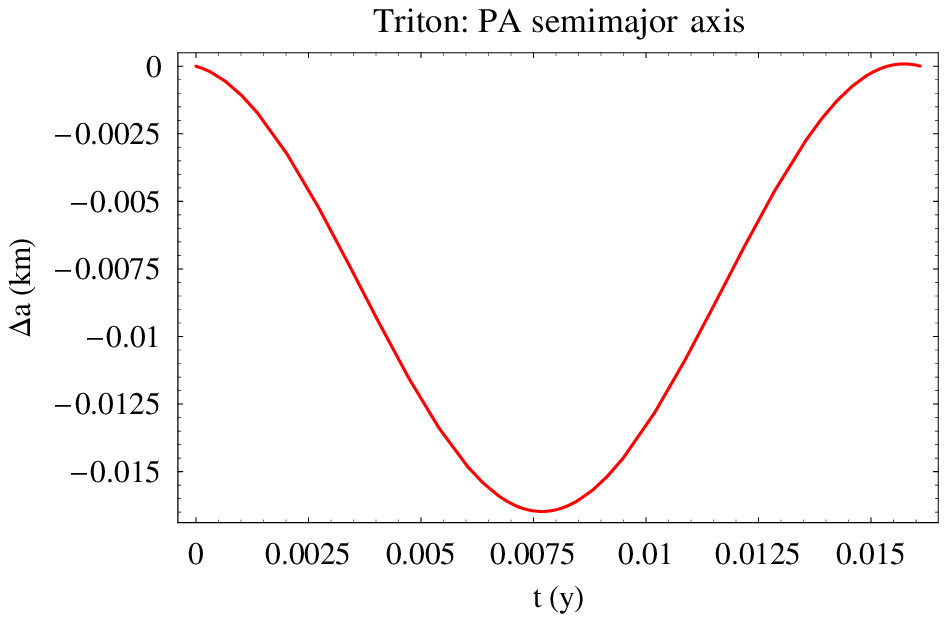}
\epsfysize= 5.5 cm\epsfbox{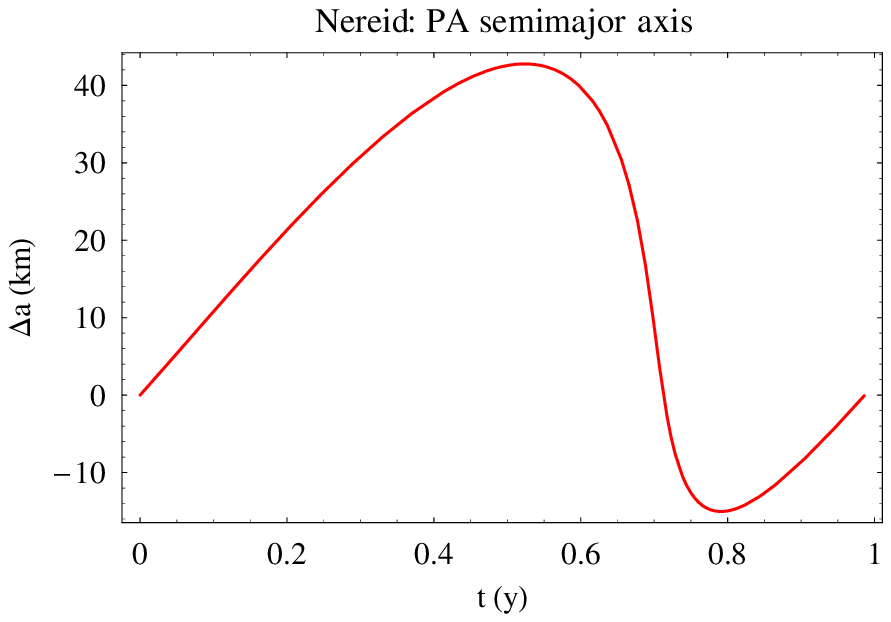}
\epsfysize= 5.5 cm\epsfbox{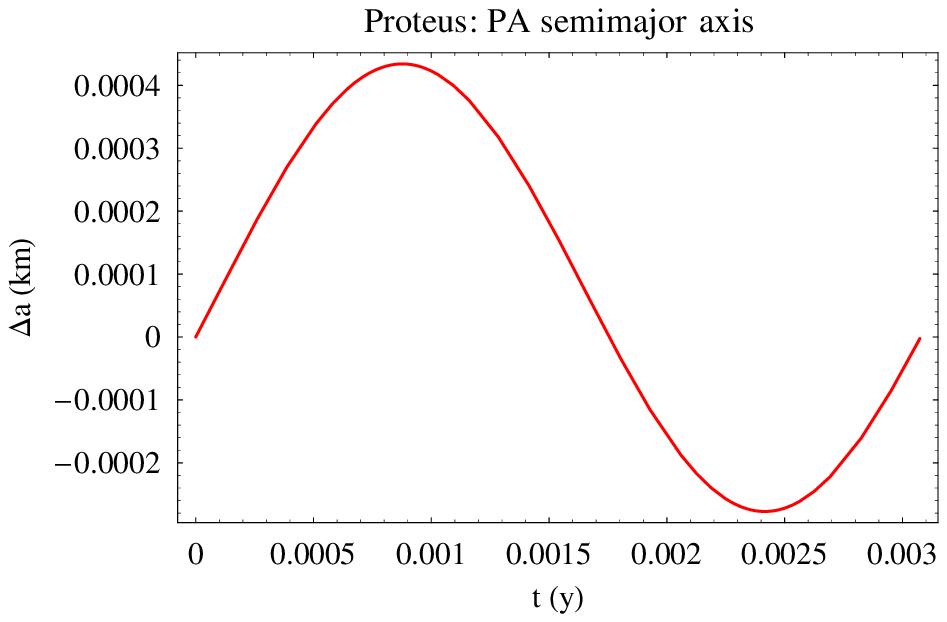}
     }
}
\caption{Pioneer-induced shifts on the semimajor axes for Triton, Nereid and Proteus. They have been obtained by numerically integrating with MATHEMATICA the equations of motions of the three satellites with and without the PA in the form $A^{\rm Pio}{\bds n}_{\odot}$ starting from the same set of initial conditions at the epoch 31 October 1989. Then, we have computed $\Delta a(t)\doteq a^{\rm Pio}(t)-a^{\rm Newton}(t)$. The gravitational attractions by Uranus, Saturn and Jupiter and the effect of the first two even zonals of the non-spherical gravitational field of Neptune have been included. Given the relatively short time interval of the integration, the positions of Uranus, Saturn, Jupiter and the Sun have been kept fixed to their values at the epoch computed with the HORIZONS software.}\label{semia}
\end{figure}
Then, for each satellite we computed $\Delta\bds r(t)\doteq\bds r^{\rm Pio}(t)-\bds r^{\rm Newton}(t)$ and projected it onto the $R-T-N$ co-moving frame of the unperturbed orbit to have $\Delta R^{\rm Pio}(t),\Delta T^{\rm Pio}(t),\Delta N^{\rm Pio}(t)$. It turns out that our numerical integrations confirm the analytical calculations: for saving space we do not show here the pictures of the numerical integrations.}
\subsection{Confrontation with the accuracy of the orbits}\lb{subsec2}
 { Although obtained differently, Figure \ref{Triton_anal}-Figure \ref{Proteus_anal} have can} be compared with Figures 4-6 by \citet{Jac09}, which yield a measure of the accuracy of the orbits of the Neptunian satellites considered. They have been obtained by fitting the observational data sets without including the PA in the dynamical force models of Triton, Nereid and Proteus, but are useful to give us an idea about a possible detection of Pioneer-type effects in their orbital dynamics.
 { More specifically, Figures 4-6 by \citet{Jac09} display the orbit uncertainties from the consider covariance mapped into the $R-T-N$ directions. In general, the solution covariance yields an optimistic measure of the orbit uncertainties because it  does not account for possible systematic or unmodeled errors. They can  occur in the dynamical force models, in the observation modeling or in the observation themselves; \citet{Jac09} believes that the dominant systematic errors mainly reside in the observations because the models adopted fit to them at their presumed accuracies. To include the effect of neglected errors, \citet{Jac09} added some \virg{consider} parameters to the estimation process. They are quantities which are not estimated, but whose uncertainty contributes to the uncertainty in the estimated parameters. The reliability of such a procedure was subsequently tested by \citet{Jac09} refitting the orbits with different data sets and comparing the consequent changes in the orbits to the uncertainties derived from the consider covariance  for the modified data sets. It turned out that the changes in the orbits were at or below the level of uncertainties. Thus, \citet{Jac09} concluded that all important errors were properly accounted for, and that the consider covariance can reliably be adopted as a realistic measure of the orbit accuracy. }

 Figure 4 by \citet{Jac09} deals with Triton. The middle panel shows the radial distance uncertainty which oscillates between $0.8$ km and $1.2$ km over a time span of one century.
 { The middle panel of Figure \ref{Triton_anal} tells us that the Pioneer-induced radial perturbation would be as large as 300 km} for Triton, so that it seems reasonable to argue that the resulting overall anomalous shift  should not have escaped from detection over a time span { of more than one century}; note that the astrometry of Triton covers 161 yr from 1847, i.e. one year after its discovery, through 2008. Even by re-scaling the radial uncertainty by a factor 10, the situation would not change.
 The upper panel of Figure 4 by \citet{Jac09} depicts the transverse uncertainty. It linearly grows reaching a level of about 150 km after 60 yr; { the upper panel of Figure \ref{Triton_anal} shows that $\Delta T^{\rm Pio}$ is larger, although not by two orders of magnitude as in the radial case.
 }
 Thus, also a Pioneer-type transverse effect may have remained undetectable with difficulty.
 { The lower panel of Figure 4 by \citet{Jac09}) displays } the out-of-plane uncertainty { which} amounts to about 50 km after 60 yr, while the anomalous PA perturbation would be orders of magnitude smaller, { as shown by the lower panel of Figure  \ref{Triton_anal}.
 }

 The orbit accuracy of Nereid is shown in Figure 5 by \citet{Jac09}. The radial distance uncertainty linearly grows up to about $1,600$ km after 60 yr (middle panel of Figure 5 by \citet{Jac09}), while { the peak-to-peak radial PA effect is as large as $\Delta R^{\rm Pio}=17,500$ km}  for Nereid whose data set covers 59 yr from its discovery in 1949 through 2008, { i.e. one order of magnitude larger}.
 %
 The upper panel of Figure 5 by \citet{Jac09} displays the transverse accuracy which linearly grows up to about $3,000$ km. According to
{ Figure \ref{Nereid_anal},} 
 $\Delta T^{\rm Pio}=35,000$ km, { i.e. about one order of magnitude larger.}
 %
 The accuracy in the out-of-plane direction is displayed in the lower panel of Figure 5 by \citet{Jac09}; it oscillates between  a few km to 60 km. { According to the lower panel of Figure \ref{Nereid_anal}
 }, the corresponding Pioneer-type out-of-plane { peak-to-peak amplitude is as large as $800$ km, more than ten times larger}. Thus, in the case of Nereid the PA perturbations over 59 yr would be { more than one order of magnitude} larger than the corresponding orbit accuracy in all the three directions.

Figure 6 by \citet{Jac09} depicts the $T-R-N$ accuracy for Proteus whose observational record is about 20 yr long. The radial distance uncertainty, shown in the middle panel Figure 6 by \citet{Jac09}, oscillates between $2.5$ km and $4.5$ km, while the { peak-to-peak amplitude of the} anomalous PA radial signal { over a similar time span is about $20-30$ km}.
%
The transverse accuracy (upper panel of Figure 6 by \citet{Jac09})  linearly grows up to 200 km, making, thus, problematic a detection of the
corresponding PA transverse shift ($\Delta T^{\rm Pio}=120$ km).
The anomalous out-of-plane effect is several orders of magnitude smaller than the corresponding orbit accuracy shown in the lower panel of Figure 6 by \citet{Jac09} which is as large as 50 km. { Thus, in the case of Proteus a PA-type radial perturbation would be about one order of magnitude larger than the corresponding orbit uncertainty, while the transverse and out-of-plane PA signatures would have been overwhelmed  by the corresponding orbit uncertainties}.

Finally, let us conclude by noting that, concerning the direct effect of the PA on the orbital motions of the outer planets, \citet{Iorio06} showed in their Table 1 that the induced anomalous perihelion precessions $\dot\varpi_{\rm Pio}$ are in the range $83.5-116.2$ arcsec cty$^{-1}$ for Uranus-Pluto. Actually, latest determinations of the corrections $\Delta\dot\varpi$ to the standard perihleion precessions by \citet{Pit010} with the EPM2008 ephemerides summarized in her Table 8 are, instead, $-3.89\pm 3.90$ arcsec cty$^{-1}$, $-4.44 \pm  5.40$ arcsec cty$^{-1}$, $2.84\pm  4.51$ arcsec cty$^{-1}$, respectively.
\section{Discussion and conclusions}\lb{conclu}
We have investigated the impact that an anomalous, constant and uniform acceleration directed towards the Sun having the same magnitude of the PA would have on the orbital dynamics of the Neptunian satellites Triton, Nereid and Proteus which move in the deep PA region of the solar system. Long data sets covering a large number of orbital revolutions are currently available for them.

We, first,  used an analytical approach which only considered the direct PA-type perturbations on the three satellites taken separately { to work out the corresponding shifts in the radial, transverse and out-of-plane orbit components}. In fact, also the indirect effects caused by the PA-affected mutual gravitational interactions among them should be, in principle, considered. Then, we  numerically integrated the equations of motion with and without an extra-PA acceleration { confirming the analytical findings. It turned out that only secular and sinusoidal signatures are present in the three orbit components; we showed that this is not in contrast with the absence of a secular effect on the semi-major axis. No quadratic terms appear in the transverse component, as, instead, it would happen if $a$ was affected by secular signatures. }

Our analysis showed that the resulting anomalous orbital effects are much larger than the realistic orbit accuracies evaluated from a recent analysis of all the available astrometric observations by one-two orders of magnitude.
{ However,} it must be stressed that our investigation should be considered preliminary.
{ Indeed, it would be necessary to refit the entire set of observations
to the corresponding predictions computed by taking the anomalous PA effect into account.
As an alternative approach, it would also be possible to fit the predicted observations without the PA to a set of simulated observations produced by including the PA. }
Our study demonstrates that such  further investigations, { which are beyond the scopes of this paper,} should be considered worth the needed effort.
%

%
\section*{Acknowledgments}
I thank R. A. Jacobson for useful material provided. I am grateful to an anonymous referee for his/her valuable comments.


\end{document}